\documentclass[sigconf, nonacm]{acmart}

\newcommand\vldbdoi{XX.XX/XXX.XX}
\newcommand\vldbpages{XXX-XXX}
\newcommand\vldbvolume{19}
\newcommand\vldbissue{1}
\newcommand\vldbyear{2026}
\newcommand\vldbauthors{Authors}
\newcommand\vldbtitle{\shorttitle} 
\newcommand\vldbavailabilityurl{URL_TO_YOUR_ARTIFACTS}
\newcommand\vldbpagestyle{plain} 
\usepackage{framed}

\usepackage{xspace}
\usepackage{listings}
\usepackage{framed}
\usepackage{multirow}
\usepackage[normalem]{ulem}
\usepackage{algorithm}
\usepackage{algpseudocode}
\usepackage{threeparttable}
\usepackage{graphicx}
\usepackage{soul}
\usepackage{textcomp}
\usepackage{xcolor}
\usepackage{multirow}
\usepackage{booktabs}
\usepackage{balance}
\usepackage{tabularx}
\usepackage{subcaption}
\usepackage{bm}
\usepackage{url}
\usepackage{makecell}
\usepackage{wasysym}
\usepackage{marvosym}
\usepackage{enumitem}
\usepackage{utfsym}
\usepackage{pifont}
\usepackage{tikz}
\usepackage{colortbl}
\usepackage{framed}

\usepackage{algpseudocode}

\newtheorem{example}{Example}

\definecolor{cadmiumgreen}{rgb}{0.0, 0.42, 0.24}
\definecolor{dropred}{rgb}{0.75, 0.22, 0.17}

\makeatletter
\DeclareRobustCommand*\cal{\@fontswitch\relax\mathcal}
\makeatother

\newcommand{\eat}[1]{}

\newcounter{mycountermodification}

\newcounter{mycounterexp}
\renewcommand{\themycounterexp}{\arabic{mycounterexp}}
\newcommand{\expnum}{%
    \refstepcounter{mycounterexp}
    \themycounterexp
}

\newcounter{mycounterfinding}
\renewcommand{\themycounterfinding}{\arabic{mycounterfinding}}
\newcommand{\findingnum}{%
    \refstepcounter{mycounterfinding}
    \themycounterfinding 
}

\colorlet{shadecolor}{gray!20}

\definecolor{shadecolor}{RGB}{220,220,220}

\newcommand{\sstab}{\rule{0pt}{8pt}\\[-2.2ex]}

\newcommand{\bi}{\begin{itemize}}
\newcommand{\ei}{\end{itemize}}

\newcommand{\be}{\begin{enumerate}}
\newcommand{\ee}{\end{enumerate}}

\newcommand{\stitle}[1]{\sstab\noindent{\bf #1}}
\newcommand{\etitle}[1]{\vspace{0.5mm}\noindent{\underline{\em #1}}}
\newcommand{\ie}{{\em i.e.,}\xspace}
\newcommand{\eg}{{\em e.g.,}\xspace}

\usepackage{amsmath} 

\newcommand{\action}[1]{%
  \ifmmode
    \text{\texttt{\ensuremath{\langle}#1\ensuremath{\rangle}}}%
  \else
    \texttt{\ensuremath{\langle}#1\ensuremath{\rangle}}%
  \fi
}

\newcommand{\actionend}[1]{%
  \ifmmode
    \text{\texttt{\ensuremath{\langle}/#1\ensuremath{\rangle}}}%
  \else
    \texttt{\ensuremath{\langle}/#1\ensuremath{\rangle}}%
  \fi
}

\DeclareMathOperator{\Prob}{Pr}

\newcommand{\sys}{\textsc{DeepPrep}\xspace}

\begin{document}
\pagestyle{\vldbpagestyle}


\title{\sys: An LLM-Powered Agentic System for \\Autonomous Data Preparation}














\author{Meihao Fan$^\dagger$,
    Ju Fan$^\dagger$,
    Yuxin Zhang$^\dagger$,
    Shaolei Zhang$^\dagger$,
    Xiaoyong Du$^\dagger$,\\
    Jie Song$^*$,
    Peng Li$^*$,
    Fuxin Jiang$^*$,
    Tieying Zhang$^*$,
    Jianjun Chen$^*$}

\affiliation{
  $^\dagger$ Renmin University of China;
  $^*$ ByteDance;
}

\email{
  {fmh1art, fanj, yuxin.zhang, zhangshaolei98, duyong}@ruc.edu.cn
}

\email{
  {jie.song, peng.li01, jiangfuxin, tieying.zhang, jianjun.chen}@bytedance.com
}

\begin{abstract}
Data preparation, which aims to transform heterogeneous and noisy raw tables into analysis-ready data, remains a major bottleneck in data science.
Recent approaches leverage large language models (LLMs) to automate data preparation from natural language specifications. However, existing LLM-powered methods either make decisions without grounding in intermediate execution results, or rely on linear interaction processes that offer limited support for revising earlier decisions.
To address these limitations, we propose \sys, an LLM-powered agentic system for autonomous data preparation. \sys constructs data preparation pipelines through iterative, execution-grounded interaction with an environment that materializes intermediate table states and returns runtime feedback. To overcome the limitations of linear interaction, \sys organizes pipeline construction with tree-based agentic reasoning, enabling structured exploration and non-local revision based on execution feedback.
To enable effective learning of such behaviors, we propose a progressive agentic training framework, together with data synthesis that supplies diverse and complex ADP tasks.
Extensive experiments show that \sys achieves data preparation accuracy comparable to strong closed-source models (\eg GPT-5) while incurring 15$\times$ lower inference cost, while establishing state-of-the-art performance among open-source baselines and generalizing effectively across diverse datasets.

\end{abstract}

\maketitle
\begingroup\small\noindent\raggedright\textbf{PVLDB Reference Format:}\\
\vldbauthors. \vldbtitle. PVLDB, \vldbvolume(\vldbissue): \vldbpages, \vldbyear.\\
\href{https://doi.org/\vldbdoi}{doi:\vldbdoi}
\endgroup
\begingroup
\renewcommand\thefootnote{}\footnote{\noindent
This work is licensed under the Creative Commons BY-NC-ND 4.0 International License. Visit \url{https://creativecommons.org/licenses/by-nc-nd/4.0/} to view a copy of this license. For any use beyond those covered by this license, obtain permission by emailing \href{mailto:info@vldb.org}{info@vldb.org}. Copyright is held by the owner/author(s). Publication rights licensed to the VLDB Endowment. \\
\raggedright Proceedings of the VLDB Endowment, Vol. \vldbvolume, No. \vldbissue\ %
ISSN 2150-8097. \\
\href{https://doi.org/\vldbdoi}{doi:\vldbdoi} \\
}\addtocounter{footnote}{-1}\endgroup

\ifdefempty{\vldbavailabilityurl}{}{
\vspace{1em}
\begingroup\small\noindent\raggedright\textbf{PVLDB Artifact Availability:}\\
The source code, data, and/or other artifacts have been made available at 
{\url{xxxx}}.
\endgroup
}

\section{Introduction}
\label{sec:intro}

\begin{figure}[t]
    \centering 
    \includegraphics[width=0.9\columnwidth]{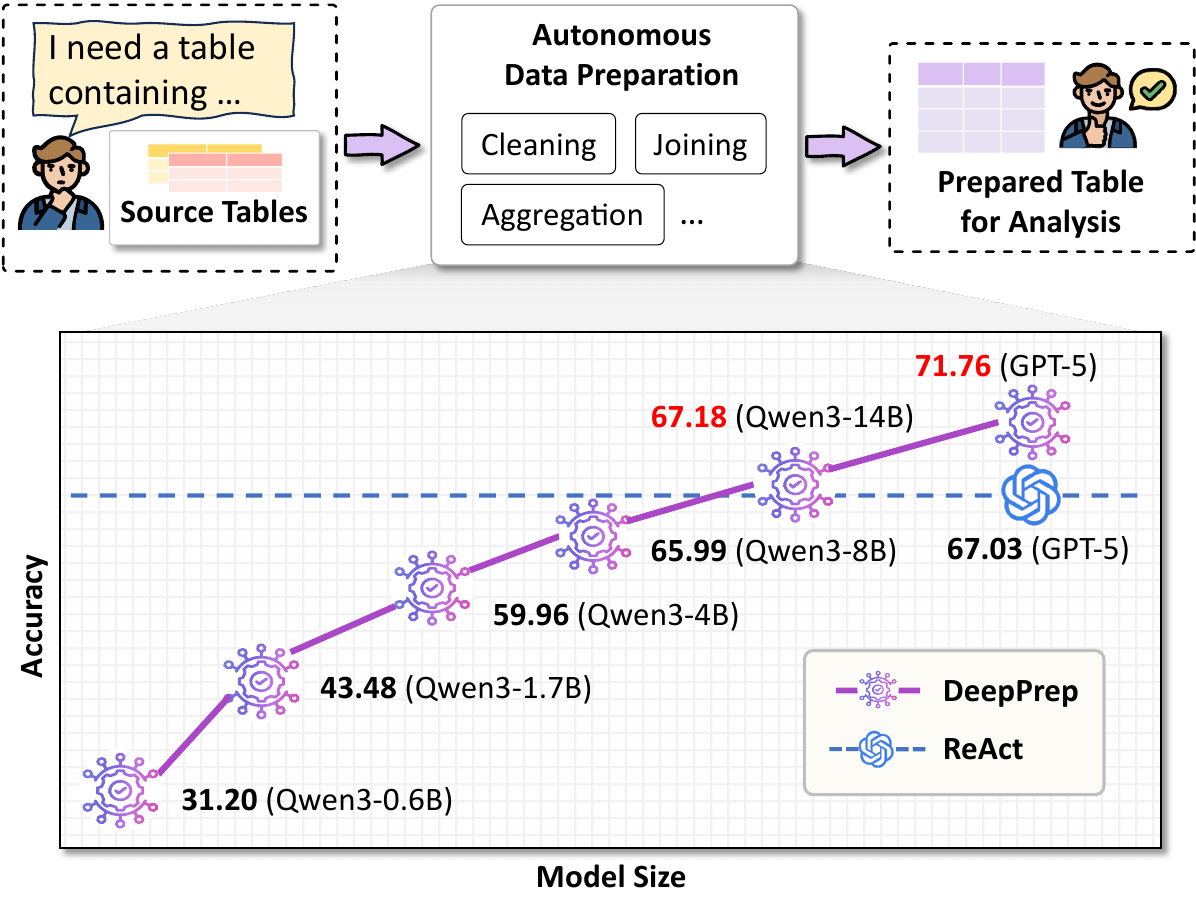}
    \vspace{-1em}
    \caption{Overview of Autonomous Data Preparation (Top) and Performance of \sys (Bottom).}
    \label{fig:introduction_experiment}
    \vspace{-1em}
\end{figure}
\begin{figure*}[!t]
    \centering 
    \includegraphics[width=0.95\textwidth]{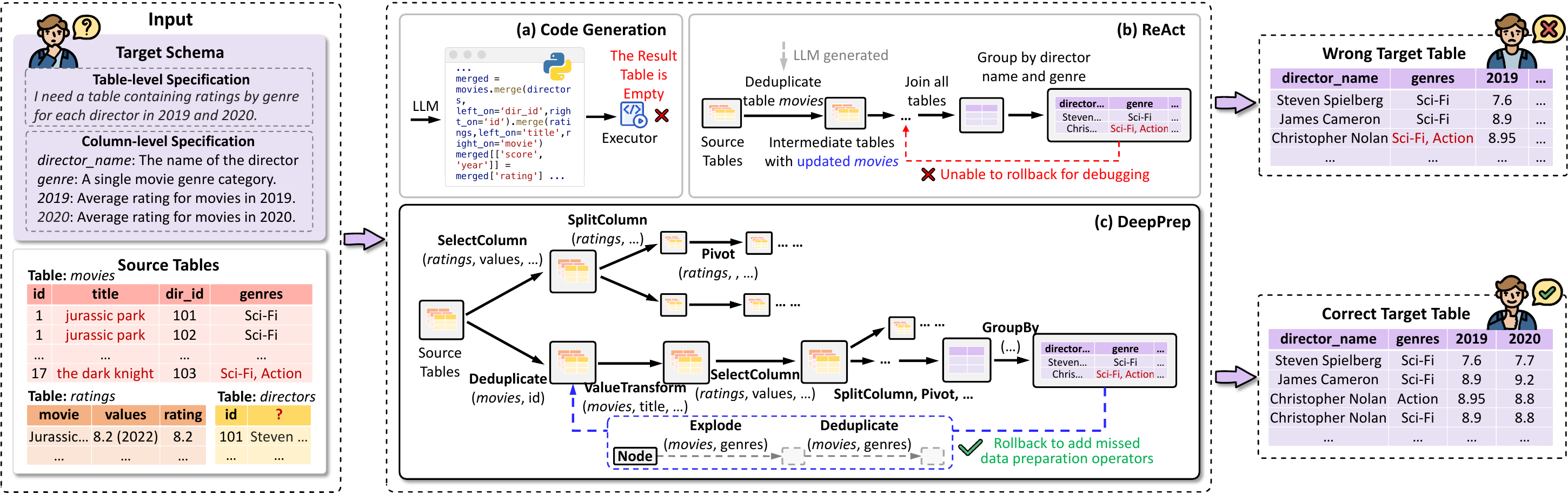}
    \vspace{-1em}
    \caption{An illustrative example highlighting the core idea of \sys.
(a) Code generation makes decisions without grounding in intermediate execution results.
(b) ReAct-style approaches follow linear reasoning with limited ability to revise earlier decisions.
(c) \sys applies a tree-based structured exploration, which enables backtracking to resolve non-local errors.
    }
    \label{fig:introduction_example}
    \vspace{-1em}
\end{figure*}

Data preparation remains a major bottleneck in data science, often consuming 60\%--80\% of end-to-end analysis time~\cite{dasu2003exploratory}. 
This is due to the heterogeneity, noise, and misalignment across sources, as well as the iterative and interdependent nature of data preparation operations such as cleaning, joining, and aggregation.
To address these challenges, autonomous data preparation (ADP) aims to automatically transform multiple source tables into analysis-ready data based on a high-level target specification, by generating an executable transformation pipeline. For example, as shown in Figure~\ref{fig:introduction_experiment}, given several source tables and a target description such as ``\emph{I need a table containing \ldots}'', the system automatically generates and executes the required transformation steps.

%

\stitle{State-of-the-Art Approaches.}
Existing approaches to autonomous data preparation broadly fall into two categories.
\begin{itemize}[leftmargin=*]
    \item Traditional methods rely on rigid specifications such as input–output examples, transformation templates, or rule-based program synthesis~\cite{barowy2015flashrelate, singh2016blinkfill, he2018transform, jin2017foofah, huang2018auto}. While effective in constrained settings, these approaches are not designed to flexibly accommodate data preparation intent expressed in natural language.



\item Recent approaches leverage large language models (LLMs) to automate data preparation from natural language specifications~\cite{dong2024large,ruan2024language,lu2025large}. Prompting-based methods generate data preparation code directly from task descriptions and examples~\cite{li2024towards, sapkota2025vibe, li2025weak}, but make decisions without systematic grounding in intermediate execution results, often leading to error-prone pipelines.
ReAct-style approaches incorporate execution feedback by interleaving reasoning with task-specific actions~\cite{yao2023react, aksitov2023rest}. However, this interaction follows a linear trajectory without explicit support for revising earlier decisions, which may be inadequate for autonomous data preparation, where early mistakes often become evident only after several dependent operations.
\end{itemize}

\stitle{Our Proposal.}
To address these limitations, we propose \textbf{\sys}, an LLM-powered agentic system for autonomous data preparation. \sys constructs data preparation pipelines through iterative, execution-grounded interaction with an environment that materializes intermediate table states and returns runtime feedback. 
Unlike one-shot code generation and linear ReAct-style interaction, which can only apply local fixes along a single trajectory, 
\sys organizes pipeline construction with \emph{tree-based agentic reasoning}.
This design enables structured exploration and \emph{non-local} revision when execution feedback indicates that earlier decisions are suboptimal.
Specifically, given source tables and a target schema, \sys operates in an iterative loop with three steps:
\begin{itemize}[leftmargin=*]
\item \textbf{Planning:} observing the current execution state and synthesizing a high-level plan for transforming existing tables toward the target schema.
\item \textbf{Orchestration:} instantiating the plan as an executable operator pipeline with specified parameters.
\item \textbf{Execution:} executing the pipeline to materialize new table states and returning runtime feedback for subsequent decisions.
\end{itemize}
Through iterative interaction with execution feedback, \sys incrementally refines its decisions and converges to a data preparation pipeline that produces a table conforming to the target schema.
\begin{example}
\label{exam:intro_tree_advantage}
Consider the example in Figure~\ref{fig:introduction_example}. The goal is to construct a table for analyzing director performance across genres using audience ratings. A linear agent following a ReAct-style interaction may prematurely join and aggregate the data, resulting in multi-valued entries such as ``Sci-Fi, Action'' in the \textit{genre} column, which violates the target schema constraint requiring single-valued categories. Correcting this error requires revisiting an earlier cleaning decision on the \textit{movies} table, \ie inserting an \texttt{Explode} operation before reapplying downstream joins and aggregations.

In contrast, \sys supports this naturally. Instead of restarting the entire process, it can backtrack to the execution state where the incorrect decision was made, expand an alternative branch with the corrected operation, and reuse valid operator prefixes from existing partial pipelines to produce a correct result.
\end{example}

\stitle{Challenges and Solutions.}
We study the research challenges posed by the above agentic framework and outline our solutions.

The first challenge is how to support structured exploration and non-local backtracking under execution feedback.
To address this, \sys maintains an explicit tree of materialized execution states during pipeline construction, where nodes represent intermediate tables produced by executed operators and edges record the applied operator pipelines.
This representation enables execution feedback to be precisely attributed to earlier decision points, allowing the agent to backtrack and revise upstream choices when downstream failures occur.
Building on this structure, \sys introduces action-structured LLM interactions that operate directly over the tree, enabling the agent to incrementally expand, revise, or backtrack branches based on execution feedback.

The second challenge is how to effectively train LLMs for the tree-based agentic reasoning.
A straightforward solution is to apply reinforcement learning with outcome-level rewards~\cite{guo2025deepseek,le2022coderl}.
However, this setting introduces two difficulties.
First, \emph{reward sparsity}: long and interdependent pipelines make outcome insufficient for learning structured exploration and non-local backtracking.
Second, \emph{data scarcity}: existing ADP datasets focus on simple, short pipelines~\cite{DBLP:journals/pvldb/LiHYWC23, lai2025auto}, limiting supervision for multi-step, execution-aware reasoning.
To address these challenges, we propose a Progressive Agentic Training framework that (i) initializes the model with basic operator usage and tree-based interaction patterns, (ii) refines the policy through multi-turn reinforcement learning with a hybrid reward that considers both outcome-level and process-level signals, and (iii) leverages data synthesis to construct diverse and complex ADP tasks for training.

\stitle{Contributions.}
We summarize our contributions as follows.

\noindent
(1) We study the problem of autonomous data preparation with support for a wide range of data preparation operators (Section~\ref{sec:preliminary}), and introduce an LLM-powered agentic framework (Section~\ref{sec:overview}).

\noindent
(2) We propose two techniques to tackle the challenges of this problem: tree-based agentic reasoning for inference (Section~\ref{sec:tree_based_agentic_reasoning}) and progressive agentic training for agent optimization (Section~\ref{sec:agentic_training_framework}).

\noindent
(3) We conduct extensive experiments on real-world ADP benchmarks.
Results show that \sys achieves accuracy comparable to closed-source models (\eg GPT-5) at {15$\times$} lower inference cost, while establishing state-of-the-art performance among open-source baselines and generalizing well across diverse datasets.

\noindent
(4) We release the source code, datasets, and a comprehensive set of trained model weights (ranging from 0.5B to 14B parameters), enabling users to deploy ADP agents under different computational budgets\footnote{We will make the materials publicly available upon institutional approval. 
}.


\section{Preliminaries}
\label{sec:preliminary}


\subsection{Problem Formulation}
\label{subsec:problem_definition}

\stitle{Source and Target Data.}
We consider a set of \emph{source tables} $\mathcal{S} = \{S_1, \ldots, S_n\}$. Each source table is denoted as $S_i = (\Sigma_i, D_i)$, where $\Sigma_i$ denotes the schema of $S_i$ and $D_i$ denotes the set of tuples conforming to $\Sigma_i$. The schema $\Sigma_i$ captures both table-level and column-level specifications and is defined as $\Sigma_i = (\tau_i, C_i)$. Here, $\tau_i$ represents a textual description that summarizes the semantic meaning of the table, and $C_i = \{c_i^{1}, \ldots, c_i^{m_i}\}$ denotes the set of column specifications. Each column specification $c_i^{j}$ includes column-level metadata such as column name, data type, and semantic description.
In practice, the schema $\Sigma_i$ of a source table may be partially specified or entirely unavailable. In such cases, table- and column-level specifications can be inferred from the raw data in $D_i$. 
For example, each source table $S_i$ in Figure~\ref{fig:introduction_example} consists of a schema $\Sigma_i$ which may be specified by table- and column-level information, and a set $D_i$ of data tuples, \eg (1,``jurassic park'', 101, ``Sci-Fi'').


Similarly, the \emph{target table} is denoted as $T=(\Sigma^{\ast}, D^{\ast})$, where $\Sigma^\ast = (\tau^\ast, C^\ast)$ specifies the target schema and $D^\ast$ denotes the target tuples to be produced. 
Note that, in autonomous data preparation, only the target schema $\Sigma^\ast$ is given as input, while the target tuples $D^\ast$ are unknown and must be generated by transforming and cleaning the source data.
For example, in Figure~\ref{fig:introduction_example}, the target schema $\Sigma^{\ast}$ is specified by a table-level description (\eg ``\emph{I need a table containing \ldots}'') and a set of column specifications, each including a column name (\eg ``\emph{director\_name}'') and a corresponding semantic description (\eg ``\emph{The name of the director.}'').

%

\stitle{Data Preparation Pipeline.}
We model the process of data preparation as a composition of \emph{operators} applied to source tables.
Specifically, an operator represents a primitive data transformation. 
%
Formally, an operator $o$ is characterized by its type $\kappa$ and parameters $\theta$, and defines a function $o_{\theta}: \mathcal{T} \rightarrow \mathcal{T}$, where $\mathcal{T}$ denotes a set of tables, and each table is of the form $(\Sigma, D)$ as defined previously. Given an input table set $\mathcal{T}_{\tt in}$, the operator produces an output table set, \ie $\mathcal{T}_{\tt out}=o_{\theta}(\mathcal{T}_{\tt in})$.
Note that different operator types capture different classes of data transformations, such as filtering, joining, aggregation, and cleaning. The current operator types supported by \sys will be introduced in Section~\ref{subsec:operator_space}.

We then define a data preparation pipeline (or pipeline for short) as an ordered sequence of operators applied iteratively to transform a set of source tables $\mathcal{S}$ into a target table $\hat{T}$. We use the symbol ``$\to$'' to denote the left-to-right application order between adjacent operators, \ie the output produced by an operator is fed as the input to the operator to its right. Formally, a pipeline $\mathcal{P}$ is defined as an ordered composition of $k$ operators over $\mathcal{S}$, \ie
\begin{equation} 
\label{eq:pipeline_definition}
    \mathcal{P} = (\mathcal{S}, o^{(1)}_{\theta_1} \to o^{(2)}_{\theta_2} \to \cdots \to o^{(k)}_{\theta_k}).
\end{equation}
Given a set of source tables $\mathcal{S}$, the execution of pipeline $\mathcal{P}$ produces a sequence of intermediate table sets, \ie 
$
    \mathcal{T}_{0} = \mathcal{S}, ~\mathcal{T}_i =  o^{(i)}_{\theta_i}(\mathcal{T}_{i-1}), i=1,\ldots,k, 
$
where $\mathcal{T}_k$ denotes the final output of the pipeline, which contains only the target table $\hat{T}$, \ie $\mathcal{T}_k=\{\hat{T}\}$.
%
Consider operator \texttt{Deduplicate (movies, [id], first)} as an example. 
Its operator type $\kappa$ is \texttt{Deduplicate} and its parameters $\theta$ refer to (\texttt{\textit{movies}, [id], first}). Executing this operator on source tables produces intermediate tables consisting of a cleaned \textit{movie}, while tables \textit{ratings} and \textit{directors} remain unchanged.

\stitle{Problem Statement.}
Given a set of source tables $\mathcal{S} = \{S_1, \ldots, S_n\}$, a target schema $\Sigma^{\ast}$, and a set of available operator types $\mathcal{O}$, the autonomous data preparation (ADP) problem is to find an optimal data preparation pipeline $\mathcal{P}$ that transforms $\mathcal{S}$ into a target table $\hat{T} = (\Sigma^{\ast}, \hat{D})$, where $\hat{D}$ denotes the tuples produced by $\mathcal{P}$ that satisfy the target schema $\Sigma^{\ast}$.
We assume a ground-truth target table $T^\ast=(\Sigma^\ast, D^\ast)$, which is used \emph{only} for evaluation. Thus, the optimality of a pipeline is defined by the quality of $\hat{D}$ with respect to $D^\ast$ under task-specific evaluation metrics, which will be introduced in the experimental setup (Section~\ref{subsec:exp-setup}).
\begin{example}

Figure~\ref{fig:introduction_example} provides an example of an ADP task with three source tables: \textit{movies}, \textit{ratings}, and \textit{directors}. The goal is to generate a target table that conforms to a given target schema. The target schema describes both the overall semantics of the table and its desired structure. Specifically, it provides a table-level summary of the intended content and specifies column-level requirements, including column names (\eg \textit{genre}), semantic meanings (\eg movie genre category), and data constraints (\eg single-valued values).
To achieve this goal, we construct a data preparation pipeline composed of multiple operators, such as \texttt{Deduplicate} and \texttt{Join}. Each operator is applied to the current set of tables and produces updated intermediate tables. For example, applying \texttt{Join} merges the \textit{movies} and \textit{directors} tables on movie titles, yielding a table named \textit{movies\_directors\_join} that integrates movie data with director names, while keeping the \textit{ratings} table unchanged.
After executing all operators in the pipeline, the process produces the target table, shown on the right of Figure~\ref{fig:introduction_example}. 
\end{example}



\subsection{Supported Data Preparation Operators}
\label{subsec:operator_space}
\sys supports a broad set of data preparation operators and is designed to be extensible, allowing new operator types to be incorporated as needed. In total, we consider 31 operators, organized into eight categories that cover the major transformation patterns in data preparation pipelines, including data cleaning, normalization, structural reshaping, table combination, and aggregation. 



\subsubsection{Data Cleaning}
Data cleaning operators improve data quality by detecting, fixing, or removing erroneous, missing, or duplicate records, and are commonly applied in data preparation pipelines.
\begin{itemize}[leftmargin=*]
    \item \texttt{DropNA(table, subset, how)} removes rows with missing values in specified columns, with \texttt{subset} specifying the columns and \texttt{how} (``any'' or ``all'') controlling the removal criterion.
    \item \texttt{MissingValueImputation(table, column, mode)} imputes missing values in a specified column using a statistical strategy, with \texttt{mode} selecting the method (\eg mean, median, or mode).
    \item \texttt{Deduplicate(table, subset, keep)} removes duplicate rows based on specified columns, retaining either the first or last occurrence as determined by \texttt{keep}.
    \item \texttt{ErrorDetection(table, column, func)} identifies invalid records in a specified column based on a user-defined function.
    \item \texttt{OutlierDetection(table, column, action)} detects statistical outliers in a specified numerical column, with \texttt{action} determining whether outliers are removed or flagged.
\end{itemize}

\subsubsection{Value Normalization}
Value normalization operators standardize field values and data types to ensure semantic consistency.
\begin{itemize}[leftmargin=*]
    \item \texttt{ValueTransform(table, column, func)} applies a user-defined transformation function to values in a specified column to standardize their format.
    \item \texttt{StandardizeDatetime(table, column, format)} converts values in a specified column to a user-defined datetime format.
    \item \texttt{CastType(table, column, dtype)} casts values in a specified column to a target data type defined by \texttt{dtype}.
\end{itemize}

\subsubsection{Schema Editing}
Schema editing operators modify table structure and column semantics, enabling column selection, renaming, derivation, and restructuring to align raw data with the target schema.
\begin{itemize}[leftmargin=*]
    \item \texttt{RenameColumn(table, rename\_map)} renames columns according to a specified mapping from original to target names.
    \item \texttt{AddNewColumn(table, name, func)} adds a new column computed from existing data using a row-wise function.
    \item \texttt{DropColumn(table, columns)} removes specified columns.
    \item \texttt{SplitColumn(table, source, target, func)} splits a column into multiple columns using a user-defined splitting function.
    \item \texttt{Concatenate(table, columns, target, func)} combines multiple columns into a single column using a formatting function.
    \item \texttt{SelectColumn(table, columns)} retains only the specified columns.
    \item \texttt{Subtitle(table, title, target\_col)} adds a new column populated with a constant title or subtitle value.
\end{itemize}

\subsubsection{Row Selection}
Row selection operators manipulate tables by filtering, ordering, or selecting subsets of rows.
\begin{itemize}[leftmargin=*]
    \item \texttt{Filter(table, func)} selects rows satisfying a given predicate.
    \item \texttt{Sort(table, by, ascending)} orders rows in a table based on specified columns, with \texttt{ascending} controlling the direction.
    \item \texttt{TopK(table, k)} retains the top-$k$ rows of the table.
\end{itemize}

\subsubsection{Aggregation}
Aggregation operators summarize data by grouping records and computing statistics, transforming row-level data into compact, analysis-ready representations.
\begin{itemize}[leftmargin=*]
    \item \texttt{GroupBy(table, by, agg)} groups rows by specified columns and applies aggregation functions defined by \texttt{agg}.
    \item \texttt{Count(table)} returns the total number of rows in the table. We treat \texttt{Count} as a primitive aggregation operator for convenience, though it can be viewed as a special case of \texttt{GroupBy} without grouping keys.
    \item \texttt{CalculateStatistic(table, stat, func)} computes a global statistic over the table using a user-defined function, returning a scalar value that summarizes a table-level property.
\end{itemize}

\subsubsection{Table Combination}
Table combination operators integrate data from multiple tables into a unified representation, enabling multi-source data preparation pipelines.
\begin{itemize}[leftmargin=*]
    \item \texttt{Join(left, right, on, how)} combines two tables based on specified key columns using a join type defined by \texttt{how}.
    \item \texttt{Union(tables, how)} vertically combines multiple tables with compatible schemas, with \texttt{how} determining whether duplicate rows are retained or removed.
    \item \texttt{Append(table, other)} appends one table to another by vertically concatenating their rows, without performing duplicate elimination as in \texttt{Union}.
\end{itemize}

\subsubsection{Table Reshaping}
Table reshaping operators reorganize the structural layout of tables, such as converting between wide and long formats or expanding nested fields.
\begin{itemize}[leftmargin=*]
    \item \texttt{Pivot(table, index, columns, values, aggfunc)} reshapes a table from long to wide format by aggregating values at the intersection of specified index and column identifiers.
    \item \texttt{Stack(table, id\_vars, value\_vars)} reshapes a table to long format by collapsing multiple columns into key–value pairs.
    \item \texttt{WideToLong(table, stubnames, i, j)} reshapes wide-format data into long format by parsing column naming patterns.
    \item \texttt{Transpose(table)} swaps the rows and columns of the table.
    \item \texttt{Explode(table, column)} expands a column containing list-valued entries into multiple rows.
\end{itemize}

\subsubsection{Program Synthesis}
The program synthesis operator provides a general mechanism for handling transformations that cannot be expressed using the above predefined operators.
\begin{itemize}[leftmargin=*]
    \item \texttt{ExeCode(tables, target, func)} generates a target table from input tables by executing Python code synthesized by an LLM, enabling transformations that cannot be expressed using the above predefined operators.
\end{itemize}

\section{An Overview of \sys}
\label{sec:overview}
\begin{figure}[t]
    \centering 
    \includegraphics[width=0.9\columnwidth]{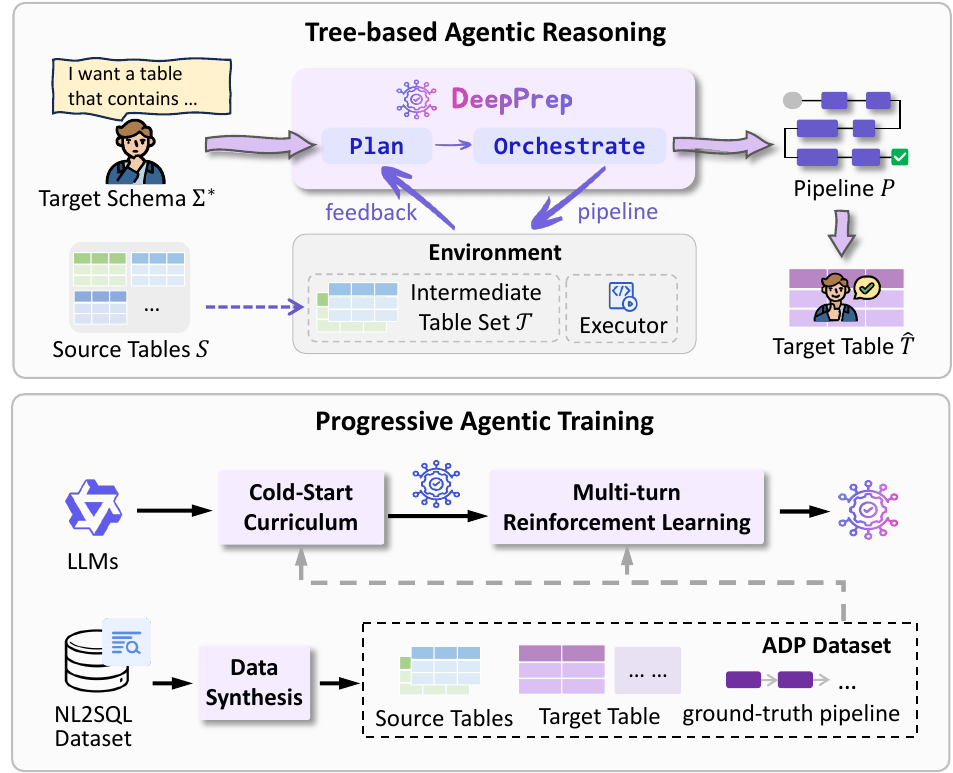}
   \vspace{-1em}
    \caption{An overview of \sys.
    }
    \label{fig:framework_overview}
   \vspace{-1em}
\end{figure}

\sys employs an agentic framework to autonomously orchestrate data preparation pipelines, as illustrated in Figure~\ref{fig:framework_overview}. Given a set of source tables $\mathcal{S}$ and a target schema $\Sigma^{\ast}$, \sys iteratively interacts with an \emph{environment} that executes data transformations and provides feedback.
The environment maintains an execution state represented by a set of intermediate tables, initialized with the source tables $\mathcal{S}$. Operator instances are executed within the environment through concrete execution backends (\eg a Python interpreter), which apply transformations to the current state and produce updated tables along with execution feedback.
Given the target schema $\Sigma^{\ast}$, \sys interacts with the environment through an iterative process that alternates among the \emph{planning}, \emph{orchestration}, and \emph{execution} steps, as described below.
\begin{itemize}[leftmargin=*]
    \item In the \emph{planning} step, the agent observes the current execution state and synthesizes a high-level textual plan that outlines how to transform the existing tables toward the target schema. 
    \item In the \emph{orchestration} step, \sys translates this plan into an executable pipeline by selecting appropriate operator types and instantiating them with specific parameters. 
    \item In the \emph{execution} step, the orchestrated pipeline is submitted to the environment, which executes the operators, updates its internal state with newly generated intermediate tables, and returns execution feedback, such as sample outputs or error traces.
\end{itemize}
Through iterative interaction with execution feedback, \sys progressively refines planning and orchestration decisions and converges to a pipeline that produces a desired target table.

\begin{example}
    Consider the ADP task shown in Figure~\ref{fig:introduction_example}. The environment is initialized with the source tables. At each iteration, \sys observes the current execution state, including serialized table contents and available schema information.
    \begin{itemize}[leftmargin=*]
    \item To align the data with the target schema, the agent first generates a high-level textual plan, such as ``Clean table \textit{movies} and then integrate it with table \textit{ratings}''. 
    \item Based on this plan, the orchestration step instantiates a pipeline, \ie a sequence of operators, \eg \texttt{Deduplicate(\ldots)} followed by \texttt{ValueTransform(\ldots)}. The orchestrated operators are then submitted to the environment for execution. 
    \item Upon execution, the environment applies the operators to the current state and returns execution feedback. Specifically, if execution is successful, the environment updates its internal state to include the cleaned movies table while retaining the original ratings and directors tables. Otherwise, execution errors are captured and propagated to the agent, which uses this feedback to revise subsequent planning and orchestration decisions.
    \end{itemize}
    The iterative process continues until the generated tables satisfy the target schema and quality requirements, at which point \sys terminates and outputs the final data preparation pipeline.
    %
\end{example}

To support the framework, we design two key components: \emph{Tree-Based Agentic Reasoning} and \emph{Progressive Agentic Training}.

\stitle{Tree-based Agentic Reasoning.}
The key challenge in agentic pipeline orchestration is supporting \emph{structured} exploration and backtracking under execution feedback. Specifically, pipeline construction involves interdependent operator decisions, where early choices may later prove suboptimal or infeasible. Linear agentic reasoning mechanisms such as ReAct~\cite{aksitov2023rest} and CoT~\cite{wang2024chain} follow a single decision trajectory and do not preserve alternative paths. Thus, once a candidate pipeline is executed, early errors can propagate to subsequent steps, and without a structured representation that retains historical states, recovery becomes difficult.

To address this challenge, we introduce a \emph{tree-based agentic reasoning} mechanism. Instead of committing to a single linear execution trajectory, \sys represents pipeline construction as a search tree, where each node corresponds to a planning state associated with a set of intermediate tables, and each edge represents applying a candidate operator pipeline. This tree structure preserves historical states and alternative construction paths, enabling systematic exploration and backtracking based on execution feedback.
Building on this tree representation, we further define LLM interaction patterns and decision logic over the tree, allowing the agent to reason about which branch to expand, revise, or abandon during pipeline construction. We detail this tree-based reasoning and interaction mechanism in Section~\ref{sec:tree_based_agentic_reasoning}.

\stitle{Progressive Agentic Training.}
The key challenge is how to effectively train LLMs for tree-based agentic reasoning.
A straightforward approach is to directly apply existing reinforcement learning methods to optimize the model using outcome-level rewards.
However, this setting poses two major difficulties.
First, \emph{reward sparsity}: data preparation pipelines often consist of long and interdependent operator sequences, and relying solely on final correctness makes it difficult for the model to learn structured exploration and non-local backtracking behaviors.
Second, \emph{data scarcity}: existing ADP datasets mainly involve simple transformations with short operator sequences~\cite{DBLP:journals/pvldb/LiHYWC23, lai2025auto}, providing insufficient supervision for learning multi-step reasoning that depends on execution feedback.

To address reward sparsity, we introduce a Cold-Start Curriculum that initializes the model with basic knowledge of data preparation operators and the tree-based agentic reasoning procedure.
Building on this initialization, we further apply multi-turn reinforcement learning with a Hybrid Reward mechanism that combines outcome-level signals with process-level feedback.
To alleviate data scarcity, we develop a data synthesis approach that generates complex ADP tasks grounded in meaningful analytical transformations, while injecting diverse data quality issues without breaking the intended transformation logic.
We detail our agentic training strategy in Section~\ref{sec:agentic_training_framework}.

\section{Tree-Based Agentic Reasoning}
\label{sec:tree_based_agentic_reasoning}


\begin{figure}[!t]
    \centering 
    \includegraphics[width=0.99\columnwidth]{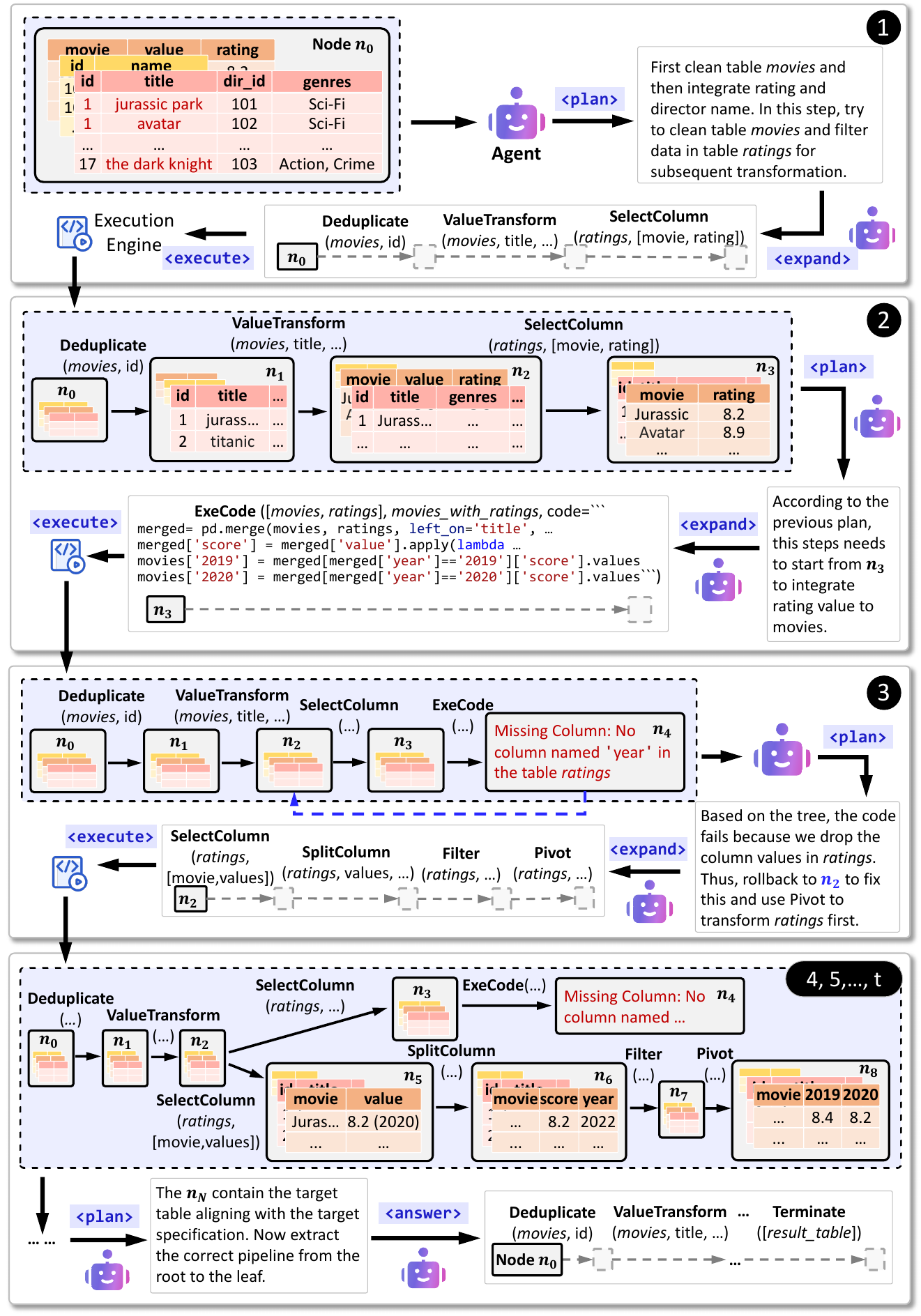}
    \vspace{-1em}
    \caption{An example of tree-based agentic reasoning.
    }
    \label{fig:agentic_reasoning_example}
     \vspace{-1em}
\end{figure}

To construct data preparation pipelines under complex dependencies and execution feedback, \sys employs a \emph{tree-based agentic reasoning} mechanism that supports structured exploration and non-local backtracking.
Specifically, we introduce an \emph{agentic reasoning tree} (Section~\ref{subsec:global_reasoning_tree}) that explicitly preserves historical states and alternative data preparation pipelines.
Building on this tree structure, we define a tree-based agentic inference procedure (Section~\ref{subsec:agentic_tree_based_definition}) through a set of explicit agentic interaction actions. These actions specify how the LLM observes the current tree state, proposes data preparation plans, and generates promising pipelines.

    
\subsection{Agentic Reasoning Tree}
\label{subsec:global_reasoning_tree}

We define the \emph{agentic reasoning tree} as a directed tree $G=(N,E)$, which represents the structured search space of pipeline construction starting from the initial source tables $\mathcal{S}$.
\begin{itemize}[leftmargin=1em]
    \item \textbf{Node ($N$).}
    Each node $n \in N$ corresponds to an \emph{environment state}, represented by the current set of intermediate tables maintained by the environment. In particular, the root node $n_0$ represents the initial state initialized with the source tables $\mathcal{S}$.
    \item \textbf{Edge ($E$).}  
    A directed edge $e=(n_i \xrightarrow{o} n_j) \in E$ represents a state transition induced by executing an \emph{operator instance} $o$ (\ie an operator type with specific parameters) orchestrated by the agent at state $n_i$. Specifically, executing $o$ updates the environment state from $n_i$ to $n_j$, yielding updated intermediate tables. 
\end{itemize}
In the agentic reasoning tree, each edge corresponds to a single operator instance, while a path from a node to another node represents an operator sequence, \ie a data preparation pipeline. 


%
%

\subsection{Tree-Based Agentic Inference}
\label{subsec:agentic_tree_based_definition}

The basic idea of our tree-based agentic inference is to develop an iterative decision-making process through a structured LLM interaction pattern.
Specifically, the LLM autonomously generates agentic \emph{interaction actions} to operate over a materialized agentic reasoning tree, incrementally expanding, revising, or backtracking tree branches based on execution feedback.

\stitle{Agentic Inference Framework.} 
Given source tables $\mathcal{S}$ and a target schema $\Sigma^\ast$, the framework produces a reasoning trajectory $R$ and finally generates an answer $A$ based on the reasoning trajectory $R$.
More formally, we model this process by a probabilistic model $\Prob(\cdot)$ over the reasoning process and the answer generation conditioned on the inputs: 
\begin{equation}
\label{eq:overall_agentic_inference}
\Prob(A,R \mid S,\Sigma^*) = 
\underbrace{\prod_{t=1}^{T} \Prob(R_t \mid R_{<t}, S, \Sigma^\ast)}_{\text{Reasoning Process}} 
 \cdot 
\underbrace{\Prob(A \mid R, S, \Sigma^*) \vphantom{\prod_{t}^{T}}}_{\text{Answer Generation}}.
\end{equation}
where $T$ denotes the length of the reasoning process and $R_t$ represents the reasoning results of the agent at step $t$.

The reasoning process is implemented by adopting an LLM-based agent to search over the agentic reasoning tree $G$ in a generative way. Specifically, the process alternates in an iteration of \textit{plan}, \textit{expand} and \textit{execute} steps where it generates logical plans for the current step based on the observation of the current tree, expands new branches and execute expanded operators to update the current tree. Based on this idea, we break the step-wise reasoning process in Eq.~(\ref{eq:overall_agentic_inference}) down into the following formula.
\begin{equation}
\label{eq:reasoning_process}
\begin{split}
&  \Prob(R_t | R_{<t}, S, \Sigma^\ast) =  \\
\underbrace{\Prob_{\text{LLM}}(\psi_t | \psi_{<t}, G_{t-1}, \Sigma^\ast)}_{\action{plan}} 
& \cdot
\underbrace{\Prob_{\text{LLM}}(\mathcal{P}_t | G_{t-1}, \psi_{t})}_{\action{expand}}
\cdot
\underbrace{\Prob_{\text{Env}}(G_t | G_{t-1}, \mathcal{P}_t)}_{\action{execute}}.
\end{split}
\end{equation}
where $\Prob_{\text{LLM}}$ denotes the generative distribution induced by the LLM agent, $\psi_t$ and $\mathcal{P}_t$ denotes the logical plan and expanded operator pipeline generated by the agent in step $t$, and $\Prob_{\text{Env}}$ represents the environment transition distribution that materializes $\mathcal{P}_t$ to get an updated agentic reasoning tree $G_t$.


\stitle{Actions for Agentic Inference.} We regulate this interaction using control tags aligned with the components in Eq.~(\ref{eq:overall_agentic_inference}) and Eq.~(\ref{eq:reasoning_process}).


\begin{itemize}[leftmargin=1em]

    \item \action{plan}.
    This action produces a high-level logical plan for the next reasoning step.
    Given the current agentic reasoning tree, the target schema $\Sigma^\ast$, and prior plans, the agent generates a plan $\psi$ that specifies the intended reasoning decision.
    The plan analyzes the current environment state to determine whether to continue expanding the tree or terminate the reasoning process. If further exploration is required, it identifies the parent node to extend and the candidate operations to apply.

    \item \action{expand}.
    This action generates a candidate operator pipeline expanded from a selected parent node $n_{\texttt{parent}}$. Conditioned on the plan $\psi$, the agent produces an operator sequence $\mathcal{P}$ to extend node $n_{\texttt{parent}}$. To ensure consistent node selection, the parent node is specified using a prefix-matching constraint: the agent must reference the full operator sequence from the root to that node. This constraint avoids ambiguous or index-based node references and enforces alignment between the proposed expansion and the recorded transformation history.


    \item \action{execute}. 
    This action applies the candidate operator pipeline through the environment and performs state transitions. Specifically, given a candidate pipeline $\mathcal{P}_t$, the environment executes its operators sequentially over the materialized tables at the selected parent node. After each operator is successfully applied, a new intermediate state node is created to store the resulting tables, and a directed edge is added from the previous state node to the new one. Repeating this process for all operators in $\mathcal{P}_t$ yields a chain of new nodes, which updates $G_{t-1}$ to $G_t$.
    If execution fails due to a runtime exception (\eg column mismatch or type error), no new state node is created; instead, the parent node is annotated with a failure state that records the error trace. The execution feedback will guide subsequent actions.
    
    \item \action{answer}.
    This action terminates the reasoning process and returns the final pipeline.
    It is triggered when a leaf node state satisfies the target schema $\Sigma^\ast$.
    The root-to-leaf operator path associated with that node is extracted as the final pipeline $\mathcal{P}^\ast$ and returned as the output.
%

\end{itemize}

\stitle{Agentic Inference Procedure.} 
We extend the LLM vocabulary with the action tags defined above, so that our tree-based reasoning can be expressed as a structured language modeling process.

During inference, the model iteratively generates tagged outputs corresponding to the defined interaction actions. Specifically, in each iteration, the model first generates a plan in the form of \action{plan}\ldots\actionend{plan}, which specifies the next reasoning decision. Based on this plan, the model predicts an operator expansion wrapped by \action{expand}\ldots\actionend{expand}. The produced pipeline is then executed by the environment, and the resulting state and execution feedback are recorded through \action{execute}\ldots\actionend{execute}. The above process iterates until a state satisfies the target schema. When this condition holds, the final pipeline is returned using \action{answer}\ldots\actionend{answer}, which terminates the reasoning process.

%
%
\begin{example}
Figure~\ref{fig:agentic_reasoning_example} illustrates an example reasoning trajectory for the task in Figure~\ref{fig:introduction_example}.
\textbf{\textcircled{1} Initialization:}
Reasoning starts from node $n_0$, which corresponds to an initial state associated with source tables $\mathcal{S}$.
During inference, the model first generates a \action{plan} action, specifying a logical plan that cleans the \textit{movies} table and select relevant columns in the \textit{ratings} table. An \action{expand} action then produces operators \texttt{Deduplicate}, \texttt{ValueTransform} and \texttt{SelectColumn}, which are applied through \action{execute} to produce new states, \ie $n_1$, $n_2$ and $n_3$.
%
\textbf{\textcircled{2} Execution Failure:}
Subsequent \action{plan} and \action{expand} actions proposes integrating rating data using an \texttt{ExeCode} operator.
However, during \action{execute}, a missing-column runtime error is raised, and the transition is recorded as a failure at $n_4$ without creating a new valid state.
%
\textbf{\textcircled{3} Backtracking and Alternative Expansion:}
Based on the recorded failure feedback, a new \action{plan} analyzes the error and tries to debug it by keeping the column values of ratings table when applying the operator \texttt{SelectColumn}. Specifically, it selects $n_2$ as the parent state for debugging and further exploration.
In addition to applying the updated \texttt{SelectColumn}, the \action{expand} action also proposes an alternative operator sequence (\texttt{SplitColumn}, \texttt{Filter}, \texttt{Pivot}). These operators are applied to generate new states ($n_5, n_6, n_7,n_8$).
%
\textbf{\textcircled{t} Termination:}
When a leaf state satisfies the target schema $\Sigma^\ast$, an \action{answer} action is issued to extract the corresponding root-to-leaf operator path as the final pipeline.
\end{example}

\section{Agentic Training Framework}
\label{sec:agentic_training_framework}



We propose a \emph{Progressive Agentic Training} framework for optimization. To mitigate reward sparsity, the framework includes a \emph{Cold-Start Curriculum} for initialization (Section~\ref{subsec:cold_start}), followed by a \emph{Multi-turn RL} stage with a \emph{Hybrid Reward} mechanism (Section~\ref{subsec:rl_reward}). To alleviate data scarcity, we develop a \emph{Data Synthesis} method (Section~\ref{subsec:data_synthesis}) that generates diverse ADP tasks for effective training.

\subsection{Cold-Start Curriculum}
\label{subsec:cold_start}

To mitigate the ``cold start'' issue where an untrained LLM fails to produce executable pipelines, we design a Cold-Start Curriculum module that initializes the LLM's reasoning capability for our structured and execution-grounded planning. 
Specifically, pre-trained LLMs generally lack an explicit understanding of data preparation operator syntax and the tree-based agentic reasoning procedure, which are previously described. Thus, without specialized supervision, the model cannot generate meaningful action sequences for ADP tasks. 
To address this, we employ a two-stage supervised fine-tuning (SFT) curriculum.

\stitle{Stage 1: Operator Syntax Learning.}
This stage focuses on the syntax and semantics of individual operators.
Given source tables and a ground-truth operator pipeline $\mathcal{P}_{\text{total}}$ as defined in Eq.~(\ref{eq:pipeline_definition}), consisting of a sequence of operators $(o^{(1)}, \dots, o^{(N)})$. We first execute these operators sequentially to collect a sequence of generated intermediate table sets $\{\mathcal{T}_{0}, \dots, \mathcal{T}_{N}\}$.
Then, we sample a contiguous sub-sequence of operators $\mathcal{P}_{\text{sub}} = (o^{(i)}, \dots, o^{(j)})$ with the corresponding input intermediate table set $\mathcal{T}_{\text{in}}=\mathcal{T}_{i-1}$ and output intermediate table set $\mathcal{T}_{\text{out}}=\mathcal{T}_{j}$.
In this way, we construct the operator syntax learning dataset $\mathcal{D}_{\text{op}}$ consisting of training triples $(\mathcal{P}_{\text{sub}}, \mathcal{T}_{\text{in}}, \mathcal{T}_{\text{out}})$.
Let $\phi(\cdot)$ denote the serialization function. It serializes $\mathcal{P}_{\text{sub}}$ by concatenating all serialized operators and intermediate table set $\mathcal{T}$ by serializing it into markdown format. We conduct SFT to optimize the following loss $\mathcal{L}_{\text{op}}$:
\begin{equation}
\label{eq:sft_op_learn}
\mathcal{L}_{\text{op}} 
= - \sum_{(\mathcal{P}_{\text{sub}}, \mathcal{T}_{\text{in}}, \mathcal{T}_{\text{out}}) \in \mathcal{D}_{\text{op}}}
\log \Prob_{\text{LLM}}(
\phi(\mathcal{P}_{\text{sub}})
\mid 
\phi(\mathcal{T}_{\text{in}}), \phi(\mathcal{T}_{\text{out}})
).
\end{equation}
Through Eq.~(\ref{eq:sft_op_learn}), the agent learns to recognize operator function signatures and argument formats, while isolating operator execution logic from higher-level planning decisions.

\stitle{Stage 2: Reasoning Procedure Learning.}
This stage aligns the model with the \textit{Agentic Tree-based Reasoning} mechanism (Section~\ref{subsec:agentic_tree_based_definition}).
To achieve this, we synthesize a dataset of high-quality reasoning trajectories by distilling knowledge from strong teacher models (\eg DeepSeek-R1).
Specifically, for each ADP task defined by source tables $\mathcal{S}$ and a target schema $\Sigma^\ast$, we aim to generate a ground-truth trajectory $\tau$.
To enhance robustness, we apply input perturbations $\Phi$, such as randomly shuffling rows and columns in the source tables.
Next, we employ In-Context Learning~\cite{dong2024survey} to guide the teacher model. By utilizing elaborate prompts that exemplify the tree-based reasoning process defined in Eq.~(\ref{eq:overall_agentic_inference}), we generate a set of candidate trajectories $\mathcal{U}$.
To unify the training objective, we represent each trajectory as a sequence $\tau = (r_1, \dots, r_m)$. Here, the initial subsequence $(r_1, \dots, r_{m-1})$ corresponds to the iterative reasoning steps $(R_1, \dots, R_T)$ defined in Eq.~(\ref{eq:reasoning_process}), while the final element $r_m$ corresponds to the answer generation $A$ in Eq.~(\ref{eq:overall_agentic_inference}).
These candidates are filtered using a process reward function $R_{\text{llm}}(\tau)$ (detailed in Section~\ref{subsec:rl_reward}), and only the top-ranked trajectories are retained to form the final dataset $\mathcal{U}^\ast$.
Finally, the model is fine-tuned on $\mathcal{U}^\ast$ to minimize the masked negative log-likelihood:
\begin{equation}
\label{eq:sft_reasoning_learn}
\mathcal{L}_{\text{reason}} 
= - \sum_{\tau \in \mathcal{U}^\ast} 
\sum_{t=1}^{m} M_t \cdot \log \Prob_{\text{LLM}}(\phi(r_t) \mid \phi(r_{<t}), \phi(\mathcal{S}), \Sigma^\ast).
\end{equation}
Here, $r_t$ encapsulates both the current tree state and the agent's generated response. The serialization function $\phi(\cdot)$ transforms these structured components into a linear token sequence.
Crucially, we introduce a loss mask $M_t$ to ensure that gradients are only calculated for tokens corresponding to the agent's outputs.



\subsection{Multi-turn RL with Hybrid Reward}
\label{subsec:rl_reward}

\stitle{Motivation.}
Data preparation pipeline generation is a sequential decision process with execution feedback, where each action affects subsequent states and future operator choices. Optimizing such behavior cannot be reduced to single-shot generation. Instead, the policy should be trained over multi-turn interaction trajectories. In each turn, the agent produces structured actions (\eg \action{plan} and \action{expand}), the environment executes the proposed operators, and feedback is returned. Then, subsequent decisions are conditioned on these execution outcomes. 

\stitle{Basic Idea.}
The above setting motivates us to develop a multi-turn reinforcement learning method, where the policy is optimized over action sequences rather than single outputs. 
Specifically, under our interaction framework, each trajectory contains both (i) agent-generated tokens (actions and their contents) and (ii) environment-generated tokens, such as serialized intermediate tables and execution traces. In particular, since environment-generated tokens are deterministic execution outputs rather than policy decisions, they are excluded from policy gradient optimization. That is, we apply a binary token mask that removes gradients on tokens inside \action{execute} blocks, and optimize the policy only over agent decision tokens within \action{plan}, \action{expand}, and \action{answer}.

We optimize the policy using Group Relative Policy Optimization (GRPO)~\cite{shao2024deepseekmath}, which estimates baselines from group-level reward statistics instead of training a separate value network, supporting more stable optimization over multi-turn trajectories. Using only final pipeline correctness as the reward, however, results in sparse and delayed learning signals. We therefore introduce a hybrid reward design, as described below.

\stitle{Hybrid Reward.}
We design a hybrid reward mechanism that provides denser learning signals by evaluating both outcome quality and reasoning behavior. The overall reward $R(\tau)$ for a trajectory $\tau$ is defined as a weighted sum of three components:
\begin{equation}
    R(\tau) = \alpha \cdot R_{\text{out}}(\tau) + \beta \cdot R_{\text{part}}(\tau) + \gamma \cdot R_{\text{llm}}(\tau).
\end{equation}
where $R_{\text{out}}$, $R_{\text{part}}$, and $R_{\text{llm}}$ correspond to strict correctness, partial progress, and reasoning process signals, respectively, and $\alpha, \beta, \gamma$ are hyperparameters that balance their contributions. 
Now, we detail each of the three components $R_{\text{out}}$, $R_{\text{part}}$, and $R_{\text{llm}}$ as follows. 


\etitle{Outcome Reward $R_{\text{out}}$.}
This reward provides a direct success signal based on final output correctness. When the \action{answer} action is issued, the generated resulting target table $\hat{T}$ is evaluated by:
\begin{equation}
R_{\text{out}}(\tau) = \mathbb{I}(\hat{T} \equiv T^*).
\end{equation}
where $\mathbb{I}(\cdot)$ denotes the indicator function.


\etitle{Partial Reward $R_{\text{part}}$.}
This reward provides a partial-progress signal when strict correctness is not achieved ($R_{\text{out}}(\tau)=0$). We measure the similarity between the generated table $\hat{T}$ and ground-truth $T^*$, and define $R_{\text{part}}(\tau) = \texttt{avg}(S_{sch}, S_{shp}, S_{cnt})$, where $S_{sch}, S_{shp}, S_{cnt}$ evaluate alignment in schema, shape and content, respectively:
\begin{equation} 
\begin{aligned}
S_{sch} &= \frac{|C_{\hat{T}} \cap C_{T^*}|}{|C_{\hat{T}} \cup C_{T^*}|}, \quad
S_{shp} = \exp\!\left(- \left| \frac{|\hat{D}| - |D^*|}{|D^*|} \right| \right), \\
S_{cnt} &= \frac{1}{|C_{m}|} \sum_{c \in C_{m}} \frac{\sum_{i} \mathbb{I}(\hat{D}[c]_i = D^*[c]_i)}{\max(|\hat{D}|, |D^*|)}.
\end{aligned}
\end{equation}

%
%
%
%

Here, $C_{\hat{T}}$ and $C_{T^\ast}$ denote the column-name sets, $|\hat{D}|$ and $|D^\ast|$ denote the row counts, and $C_{m} = C_{\hat{T}} \cap C_{T^*}$ is the set of matched columns used for cell-wise value comparison.
%

\etitle{Process Reward $R_{\text{llm}}$}.
Relying solely on data similarity rewards may lead to \emph{reward hacking}, which has been studied in other reinforcement learning settings~\cite{taylor2025school,zhang2025reward}, and can also arise in ADP scenarios (\eg renaming columns without performing the required data cleaning).
To discourage such ``shortcut'' strategies and promote semantically consistent decision-making, we introduce a process reward that evaluates the quality of the reasoning trajectory. We use an instruction-tuned LLM (\eg GPT-4o) as a judge to score each trajectory based on the actions defined in Section~\ref{sec:tree_based_agentic_reasoning}. 

Specifically, the judge evaluates three criteria:
(1) \textit{Plan–action consistency}, measuring whether generated pipeline (from \action{expand}) correctly implements the plan (from \action{plan}).
(2) \textit{Feedback responsiveness}, checking whether subsequent \action{plan} outputs explicitly respond to the prior execution feedback (\eg reported error traces).
(3) \textit{Backtracking justification}, verifying that parent-node switches are supported by recorded failure evidence from the current branch.


\subsection{Data Synthesis for Training}
\label{subsec:data_synthesis}
To train our agentic model, we require supervision in the form of executable data preparation pipelines together with their source and target tables. Since such ADP task instances are not available at scale, we construct them through synthesis by converting SQL benchmarks into ADP tasks consisting of source tables, a target table, and a corresponding transformation pipeline.
Synthesizing such training tasks has two key challenges. First, the generated pipelines should correspond to meaningful analytical transformations rather than arbitrary operator chains, so that the supervision reflects realistic operator dependencies. Second, the synthesized source tables should include diverse data quality issues while preserving the intended transformation logic. Excessive noise injection may break key fields or relationships~\cite{rekatsinas2017holoclean}, making it difficult to construct any feasible pipeline that produces the target table.

\stitle{ADP Task Construction from NL2SQL.}
To obtain realistic ADP tasks, we instantiate them from NL2SQL benchmarks~\cite{spider_dataset,bird_dataset}, which provide real databases paired with analytical SQL queries. For each ground-truth query $q$, we execute it to produce the target table $T^*$ and use an LLM to generate the corresponding target schema specification. We also translate $q$ into an operator pipeline by generating candidate pipelines with an LLM and selecting the shortest one that exactly reproduces $T^*$ through execution. This yields a clean task instance consisting of source tables, a target schema specification, a target table, and a task pipeline.

\stitle{Reversible Noise Injection.}
Since NL2SQL data is typically clean, we introduce controlled noise to increase data diversity for training. We inject noise by applying inverse transformations of data-cleaning operators, so that each corruption corresponds to a valid cleaning step. Specifically, we sample an operator and use an LLM to generate its inverse transformation logic. For example, the inverse of a date-standardization operator converts a normalized date (\eg ``2023-01-01'') into heterogeneous formats (\eg ``01/01/23''). After each corruption step, we verify executability by applying the corresponding cleaning operator and checking that the previous table state is restored. Only reversible corruptions are kept. Repeating this process produces dirty source tables together with a matching cleaning pipeline. The final ground-truth pipeline is formed by concatenating the cleaning pipeline with the task pipeline.

\section{Experiments}
\label{sec:experiment}



\begin{table}[t]
  \centering
  \caption{\bf Statistics of datasets.}
  \vspace{-1em}
  \label{tbl:datasets}
  \resizebox{0.99\columnwidth}{!}{
    \begin{tabular}{|c||c|c|c|c|}
      \hline
      \multirow{2}{*}{\textbf{Dataset}} & \multicolumn{2}{c|}{\textbf{Statistics}} & \multirow{2}{*}{\textbf{Pipeline Length}} & \multirow{2}{*}{\textbf{\# Op Types}} \\ \cline{2-3}
       & \textbf{\# Train} & \textbf{\# Test} & & \\
      \hline \hline
      
      \textbf{Synth-Spider} 
      & 6,788 & 2,008 & 1$\sim$28 & 31 \\ \hline
      
      \textbf{Synth-Bird} 
      & - & 1,150 & 2$\sim$25 & 31 \\ \hline
      
      \textbf{Parrot} ~\cite{ge2025text}
      & 13,965 & 1,365 & 1$\sim$8 & 17 \\
      \hline
    \end{tabular}
  }
  \vspace{-1em}
\end{table}

\subsection{Experimental Setup} 
\label{subsec:exp-setup}

\stitle{Datasets.}
We evaluate \sys on three datasets, including one public ADP benchmark and two datasets synthesized using our synthesis method.
Dataset statistics are summarized in Table~\ref{tbl:datasets}.

(1) \textbf{Parrot}~\cite{ge2025text} is a benchmark for translating natural language into data preparation pipelines. However, it does not provide ADP-style task specifications such as target schema descriptions. To make it suitable for ADP evaluation, we augment each case by generating the corresponding task specification from the target tables and operator sequences using an LLM~\cite{doubaoseed}. The resulting dataset contains 15,330 cases with pipeline lengths ranging from 1 to 8.

Note that the public benchmark \textbf{Parrot}~\cite{ge2025text} mainly contains short operator pipelines and limited operator-type coverage. We therefore synthesize two datasets to provide more complex pipelines and broader operator coverage, using our method in Section~\ref{subsec:data_synthesis}.

(2) \textbf{Synth-Spider} is a dataset synthesized by our method from the Spider NL2SQL benchmark~\cite{spider_dataset}. Using the ADP task synthesis pipeline described in Section~\ref{subsec:data_synthesis}, we convert NL2SQL cases into ADP tasks with executable operator pipelines. The resulting dataset contains 6,788 training cases and 2,008 test cases, with ground-truth pipeline lengths ranging from 1 to 28.

(3) \textbf{Synth-Bird} is another synthesized dataset constructed from the Bird benchmark~\cite{bird_dataset}, which features larger databases and more complex analytical queries. We apply the same synthesis procedure to generate ADP tasks from the Bird validation set, producing 1,150 cases with pipeline lengths ranging from 2 to 25. As we use Synth-Bird only for evaluation, this dataset only has test split. Compared to Synth-Spider, this dataset involves longer transformation chains and heavier data processing.

\stitle{Evaluation Methodology.}
For comprehensive evaluation, all models are trained only on the \textbf{Synth-Spider} training set and evaluated under two test settings: \emph{in-domain} and \emph{out-of-domain}.

\etitle{(1) In-domain Test.}
We evaluate on the \textbf{Synth-Spider} test set, which is synthesized using the same procedure as the training data but with disjoint queries and databases.

\etitle{(2) Out-of-domain Test.}
We evaluate on two datasets constructed from different sources and task formats. \textbf{Synth-Bird} is closer to the training distribution since it is generated by the same synthesis pipeline but from a different benchmark. \textbf{Parrot} is an independent dataset with different task characteristics.


\stitle{Baselines.} 
We compare \sys with both \emph{prompting baselines} and \emph{training baselines}. Prompting baselines use off-the-shelf LLMs with various inference strategies, while training baselines involve parameter updates via supervised or reinforcement learning.


\etitle{(1) Prompting Baselines.} These methods use frozen LLMs with different planning and search strategies to generate pipelines.
\begin{itemize}[leftmargin=*]
    \item {CodeGen}~\cite{li2024towards,sapkota2025vibe} prompts an LLM to generate data preparation code from the source tables and target schema.
    \item {Plan-and-Solve (PaS)}~\cite{wang2024chain} prompts an LLM using a two-stage strategy that first produces a high-level plan and then generates a complete operator sequence to obtain the target table.
    \item {ReAct}~\cite{aksitov2023rest} is a linear reasoning agent that iteratively predicts the next operator based on previous execution results.
    \item {MCTS-OP} applies Monte Carlo Tree Search, using local node expansion and scalar rewards to guide search. We follow the framework of~\cite{li2025alphasql} and keep parameters of its implementation.
    \item {MontePrep}~\cite{ge2025monteprep} applies Monte Carlo Tree Search and performs tree search over actions such as schema mapping, operator discovery, code synthesis, and refinement instead of individual operators. We use the authors’ released code with default settings.
    \item {DeepAnalyze}~\cite{zhang2025deepanalyze} is an agentic LLM that generates data processing programs using supervised and reinforcement learning.
    \item {AutoPrep}~\cite{lai2025auto} is a multi-agent framework for table reconstruction and column transformation. It generates a high-level plan and then produces data preparation programs through specialized agents. Since it does not cover as many operators as our setting, we extend it following its original design.
\end{itemize}

\etitle{(2) Training Baselines.} These baselines use the same backbone models as \sys but adopt different training strategies, allowing us to evaluate the effect of our agentic training framework.
\begin{itemize}[leftmargin=*]
    \item {Imitation Learning (IL)}~\cite{li2025llms} fine-tunes the model on the high-quality reasoning trajectories distilled from strong teacher models (\eg DeepSeek-R1). Unlike standard SFT on (input, output) pairs, this baseline learns the step-by-step reasoning process.
    \item {RL-GRPO}~\cite{shao2024deepseekmath} applies reinforcement learning with Group Relative Policy Optimization using only outcome-level rewards, without our curriculum and hybrid reward design.
\end{itemize}

%
%

\stitle{Evaluation Metrics.}
We evaluate all methods using three metrics.

\etitle{(1) Exact Match Accuracy (Acc.)} measures the percentage of test cases where the produced table $\hat{T}$ matches the ground-truth table $T^*$. We follow the table matching method in~\cite{ge2025text}, which is invariant to row and column permutations but requires exact cell-value equality.

\etitle{(2) Completion Rate (Comp.)} measures the percentage of test cases that produce a final output table. A case is counted as incomplete if execution exceeds the maximum interaction limit, triggers runtime errors, or produces an empty result.

\etitle{(3) Inference Cost (Cost)} measures the average monetary cost per case. For closed-source LLMs, we compute API cost using official pricing with caching enabled. For open-source models, we measure per-case end-to-end runtime under parallel requests and convert it to cost using the hourly price of the GPU instance. Specifically, the cost of case $i$ is computed as
$c_i = p_{\text{gpu}} \cdot t_i / 3600$,
and we report the average over all cases. We use an A800-80G GPU priced at \$0.91/hour ({\url{https://cloud.vast.ai/}).

\stitle{Backbone LLMs and \sys Variants.}
We implement \sys on both closed-source and open-source LLM backbones.

\etitle{(1) Closed-source backbones.}
We implement \sys with our proposed tree-based agentic inference directly on the most strong proprietary models, including GPT-5 (\ie gpt-5-mini-2025-08-07) and Claude-4 (\ie claude-sonnet-4-20250514).

\etitle{(2) Open-source backbones.}
We implement \sys based on open-source LLMs, \ie Qwen2.5~\cite{hui2024qwen2} (with {0.5B}, {1.5B}, {3B}, {7B} and 14B parameters) and Qwen3~\cite{yang2025qwen3} (with {0.6B}, {1.7B}, {4B}, {8B} and 14B parameters). For these models, we apply our agentic training procedure and then evaluate them with the same inference framework.

%


\stitle{Experiment Settings.}
For all prompting methods, the temperature is set to 0.01. The maximum exploration turns of \sys are set to 5. 
For all training methods, we use the AdamW optimizer with a learning rate of $6\times10^{-5}$ for SFT and $1\times10^{-6}$ for RL. 
All experiments are conducted on 16 NVIDIA A800-80G GPUs.
Detailed prompts and hyperparameter configurations are provided in~\cite{technical_report}.

%

\subsection{Overall Comparison}
\label{expsec:overall_comparison}

\begin{table}[t]
    \centering
    \caption{\bf 
    Comparison of \sys with baselines on open-source LLMs. The best and second-best results are highlighted in \textbf{bold} and \underline{underlined}, respectively.
    }
    \label{tbl:opensource_exp}
    \vspace{-1em}

    \begin{subtable}[t]{\columnwidth}
        \centering
        \caption{\bf{Results on Qwen3-14B.}}
        \label{subtbl:open_qwen14b}
        \vspace{-0.15em}
        \resizebox{0.99\linewidth}{!}{
        \begin{tabular}{|c||c|c||c|c|c|c|}
            \hline
            \multirow{2}{*}{\textbf{Method}} & \multicolumn{2}{c||}{\textbf{Synth-Spider}} & \multicolumn{2}{c|}{\textbf{Synth-Bird}} & \multicolumn{2}{c|}{\textbf{Parrot}} \\ \cline{2-7}
             & \textbf{Acc.} & \textbf{Comp.} & \textbf{Acc.} & \textbf{Comp.} & \textbf{Acc.} & \textbf{Comp.} \\
            \hline \hline
            \multicolumn{7}{|c|}{\cellcolor{gray!10}\textit{Prompting Baselines}} \\
            PaS & 7.67 & 20.67\% & 3.09 & 21.86\% & 9.56 & 22.36\% \\
            MCTS-OP & 21.91 & 72.56\% & 7.22 & {78.69\%} & 24.95 & 83.02\% \\
            MontePrep & 9.46 & 19.62\% & 18.63 & 38.15\% & 26.94 & 57.87\% \\
            AutoPrep & 36.75 & 71.61\% & 10.41 & 39.89\% & 28.54 & 74.05\% \\
            CodeGen & 45.47 & 68.18\% & 29.48 & 49.20\% & 30.83 & 81.08\% \\
            ReAct & 40.39 & 69.02\% & 16.04 & 46.56\% & 30.80 & 76.44\% \\
            \hline \hline
            \multicolumn{7}{|c|}{\cellcolor{gray!10}\textit{Training Methods (Only Trained on Synth-Spider)}} \\
            IL & \underline{61.70} & \underline{85.26\%} & \underline{48.01} & \underline{81.23\%} & \underline{37.95} & \underline{89.39\%} \\
            RL-GRPO & 51.22 & 77.68\% & 32.25 & 67.22\% & 35.06 & 82.9\% \\ \hline
            \begin{tabular}{@{}c@{}}\textbf{\sys}\\ \textbf{(Ours)}\end{tabular} & \textbf{67.18} & \textbf{97.21\%} & \textbf{54.09} & \textbf{92.26\%} & \textbf{40.44} & \textbf{96.41\%} \\
            \hline
        \end{tabular}
        }
    \end{subtable}

    \vspace{0.5em}
    
    \begin{subtable}[t]{\columnwidth}
        \centering
        \caption{\bf{Results on Qwen3-8B.}}
        \label{subtbl:open_qwen8b}
        \vspace{-0.15em}
        \resizebox{0.99\linewidth}{!}{
        \begin{tabular}{|c||c|c||c|c|c|c|}
            \hline
            \multirow{2}{*}{\textbf{Method}} & \multicolumn{2}{c||}{\textbf{Synth-Spider}} & \multicolumn{2}{c|}{\textbf{Synth-Bird}} & \multicolumn{2}{c|}{\textbf{Parrot}} \\ \cline{2-7}
             & \textbf{Acc.} & \textbf{Comp.} & \textbf{Acc.} & \textbf{Comp.} & \textbf{Acc.} & \textbf{Comp.} \\
            \hline \hline
            \multicolumn{7}{|c|}{\cellcolor{gray!10}\textit{Prompting Baselines}} \\
            PaS & 4.68 & 7.42\% & 0.65 & 4.33\% & 3.09 & 8.32\% \\
            MCTS-OP & 12.90 & 15.19\% & 0.90 & 73.56\% & 17.33 & 23.71\% \\
            MontePrep & 8.81 & 19.97\% & 13.70 & 21.96\% & 25.95 & 56.08\% \\
            AutoPrep & 26.44 & 50.85\% & 8.52 & 42.28\% & 25.20 & 70.42\% \\
            CodeGen & 6.27 & 12.10\% & 7.47 & 15.69\% & 7.27 & 25.50\% \\
            ReAct & 29.63 & 52.59\% & 10.41 & 46.22\% & 22.11 & 57.97\% \\
            DeepAnalyze & 35.66 & 82.02\% & 1.74 & 62.17\% & 29.01 & 66.67\% \\
            \hline \hline
            \multicolumn{7}{|c|}{\cellcolor{gray!10}\textit{Training Methods (Only Trained on Synth-Spider)}} \\
            IL & \underline{59.65} & \underline{90.54\%} & \underline{43.30} & \underline{80.43\%} & \underline{36.15} & \underline{91.08\%} \\
            RL-GRPO & 47.81 & 76.54\% & 29.95 & 68.69\% & 32.82 & 85.05\% \\ \hline
            \begin{tabular}{@{}c@{}}\textbf{\sys}\\ \textbf{(Ours)}\end{tabular} & \textbf{65.99} & \textbf{97.46\%} & \textbf{53.39} & \textbf{92.78\%} & \textbf{39.93} & \textbf{98.46\%} \\
            \hline
        \end{tabular}
        }
    \end{subtable}
\end{table}


\begin{figure*}[t]
    \centering 
    \includegraphics[width=0.99\textwidth]{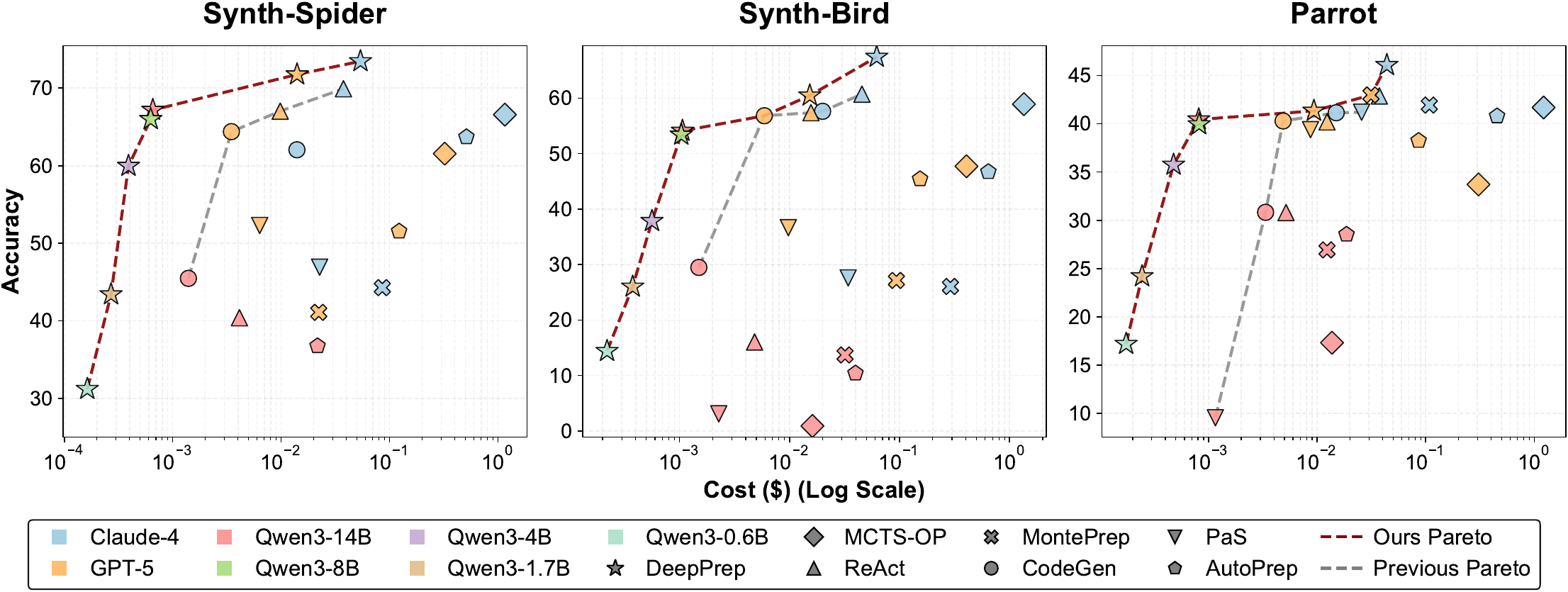}
    \caption{Cost–Accuracy trade-off comparison on Synth-Spider, Synth-Bird, and Parrot datasets. Colors represent different LLM backbones, and shapes denote different methods. The red and grey dashed lines indicate the Pareto frontiers of our method and previous baselines, respectively.
    }
    \label{fig:combined_pareto_front_compare}
  \vspace{-0.5em}
\end{figure*}

\stitle{Exp-\expnum\label{exp:training_comparison}: Accuracy and Completion Rate Comparison.}
We compare \sys with prompting baselines and training baselines on accuracy and completion rate. 
Results on Qwen3-14B and Qwen3-8B are reported in Table~\ref{tbl:opensource_exp}, as they are strong open-source backbones at two model scales commonly used in code and data tasks. Additional backbone results are presented in later experiments.

\etitle{(1) In-Domain Test.}
\sys achieves the highest accuracy and completion rate across both backbones on the Synth-Spider test set. On Qwen3-14B, it reaches 67.18 accuracy (Table~\ref{subtbl:open_qwen14b}), exceeding the strongest prompting baseline CodeGen by 21.71 points and the strongest training baseline by 5.48 points, with similar gains in completion rate. In contrast, the general-purpose data agent DeepAnalyze achieves 35.66 accuracy on the same dataset. These gains stem from combining tree-structured exploration with execution-grounded actions, enabling the agent to revise earlier decisions and avoid invalid operator sequences during pipeline construction.

\etitle{(2) Out-of-Domain Test.}
Under the single-source training method, \sys maintains a clear advantage on out-of-domain datasets. On Synth-Bird, \sys (Qwen3-14B) achieves 54.09 accuracy compared with 48.01 for IL, and the 8B model achieves 53.39 versus 43.30. On Parrot, we observe a similar ordering among methods. DeepAnalyze shows much lower performance on Synth-Bird (1.74 accuracy), indicating limited robustness when pipeline length and operator dependencies increase. These results align with our training design, where progressive curriculum and hybrid rewards expose the model to multi-step, feedback-driven reasoning trajectories rather than single-pass program generation.

\begin{shaded}
\noindent \textbf{Finding \findingnum: \sys achieves the highest accuracy and completion rate among the compared methods and shows good generalization across datasets.}
\end{shaded}

\noindent
\textbf{Exp-\expnum\label{exp:prompting_comparison}: Cost–Accuracy Trade-off Comparison.}
We compare methods under a cost–accuracy trade-off by jointly evaluating accuracy and cost. Figure~\ref{fig:combined_pareto_front_compare} plots prompting baselines and \sys variants across different LLM backbones, and shows two Pareto frontiers: \textit{Previous Pareto}, computed from baselines only, and \textit{Ours Pareto}, computed after adding \sys.


First, \sys lies on or near the Pareto frontier in most settings and extends the frontier across all three datasets. In Figure~\ref{fig:combined_pareto_front_compare}, the \textit{Ours Pareto} curve is shifted toward the upper-left relative to the \textit{Previous Pareto}, meaning that \sys achieves higher accuracy at the same cost, or similar accuracy at lower cost. This appears on Synth-Spider, Synth-Bird, and Parrot, indicating that the cost–accuracy advantage is consistent across datasets and backbones.

Second, \sys reaches accuracy close to strong closed-source baselines at substantially lower cost. On Synth-Spider, \sys (Qwen3-14B) achieves 67.18 accuracy with a very low cost, achieving a comparable accuracy with ReAct on Claude-4 (69.92 accuracy) and GPT-5 (67.03 accuracy) while reducing cost by about 57$\times$ and 15$\times$, respectively. This is attributed to our design that uses structured tree-based reasoning to guide smaller or open models toward valid pipelines with fewer failed executions.


\begin{shaded}
\noindent \textbf{Finding \findingnum: \sys extends the cost–accuracy Pareto frontier, achieving accuracy close to strong closed-source baselines at substantially lower inference cost.}
\end{shaded}
\subsection{Ablation Studies}
\label{expsec: ablation_on_policy_learning}

\begin{table}[t]
    \centering
    \caption{\bf Ablation study (Qwen3-8B).}
    \vspace{-1em}
    \label{tbl:ablation_results_split}
    
    \begin{subtable}[t]{\columnwidth}
        \centering
        \caption{\bf{Ablation study on Cold Start stage. OSL represent Operator syntax Learning module in Section~\ref{subsec:cold_start}}}
        \label{subtbl:cold_start}
        \vspace{-0.15em}
        \resizebox{0.99\linewidth}{!}{
        \begin{tabular}{|c||c|c|c|c|c|c|}
            \hline
            \multirow{2}{*}{\textbf{Method}} & \multicolumn{2}{c|}{\textbf{Synth-Spider}} & \multicolumn{2}{c|}{\textbf{Synth-Bird}} & \multicolumn{2}{c|}{\textbf{Parrot}} \\ \cline{2-7}
             & \textbf{Acc.} & \textbf{Comp.} & \textbf{Acc.} & \textbf{Comp.} & \textbf{Acc.} & \textbf{Comp.} \\
            \hline \hline
            Cold-Start & \textbf{60.66} & \textbf{91.63\%} & \textbf{45.27} & \textbf{81.39\%} & \textbf{36.22} & \textbf{92.16\%} \\
            w/o OSL & {59.65} & {90.54\%} & {43.30} & {80.43\%} & {36.15} & {91.08\%} \\
            \hline
        \end{tabular}
        }
    \end{subtable}
    
    \vspace{0.30em}
    
    \begin{subtable}[t]{\columnwidth}
        \centering
        \caption{\bf{Ablation study on the reward of Multiturn RL stage.}}
        \label{subtbl:rl_stage}
        \vspace{-0.15em}
        \resizebox{0.99\linewidth}{!}{
        \begin{tabular}{|c||c|c|c|c|c|c|}
            \hline
            \multirow{2}{*}{\textbf{Method}} & \multicolumn{2}{c|}{\textbf{Synth-Spider}} & \multicolumn{2}{c|}{\textbf{Synth-Bird}} & \multicolumn{2}{c|}{\textbf{Parrot}} \\ \cline{2-7}
             & \textbf{Acc.} & \textbf{Comp.} & \textbf{Acc.} & \textbf{Comp.} & \textbf{Acc.} & \textbf{Comp.} \\
            \hline \hline
            \sys & \textbf{65.99} & \underline{97.46\%} & \textbf{53.39} & \underline{92.78\%} & \textbf{39.93} & \underline{98.46\%} \\
            w/o $R_{\text{llm}}$ & \underline{64.54} & 96.96\% & \underline{51.67} & 92.31\% & \underline{39.12} & 94.21\% \\
            w/o $R_{\text{llm}}, R_{\text{part}}$ & 61.85 & \textbf{98.71\%} & 47.15 & \textbf{95.57\%} & 37.22 & \textbf{99.34\%} \\
            \hline
        \end{tabular}
        }
    \end{subtable}
    
  \vspace{-0.5em}
\end{table}

\stitle{Exp-\expnum\label{exp:ablation_analysis}: Ablation Studies.}
We evaluate the contribution of specific components within our agentic training framework. The results are detailed in Table~\ref{tbl:ablation_results_split}. 

Table~\ref{subtbl:cold_start} demonstrates that cold-start curriculum is useful. Removing this component causes a simultaneous decline in both accuracy and completion rate across all datasets. For instance, the completion rate on Synth-Spider drops from 91.63 to 90.54.
This degradation occurs because this component familiarizes the model with the syntax and parameter constraints of the operators. Without this alignment, the model tends to generate non-executable actions, and thus fails to complete the reasoning trajectory. 


Table~\ref{subtbl:rl_stage} validates the design of our reward function in RL.
First, removing process reward ($R_{\text{llm}}$) results in an accuracy drop, which shows that $R_{\text{llm}}$ prevents ``reward hacking''. Without semantic verification, the policy learns to satisfy superficial schema constraints, such as renaming columns, while neglecting the actual data transformation logic.
Second, relying solely on outcome rewards (w/o $R_{\text{out}}, R_{\text{part}}$) yields the highest completion rate but the lowest accuracy.
This suggests that sparse rewards induce conservative behavior. The agent avoids complex exploration or backtracking to minimize the risk of runtime errors. Thus, it prioritizes safe but incorrect paths to ensure the episode terminates successfully.


\begin{shaded}
\noindent \textbf{Finding \findingnum: The key components of the Progressive Agentic Training framework each contribute to performance, with ablations showing consistent drops when the cold-start or hybrid-reward stages are removed.
}
\end{shaded}
\subsection{Effectiveness of Training Data Synthesis}
\label{subsec:data_effectiveness}

\begin{figure}[t]
    \centering
    \includegraphics[width=0.99\columnwidth]{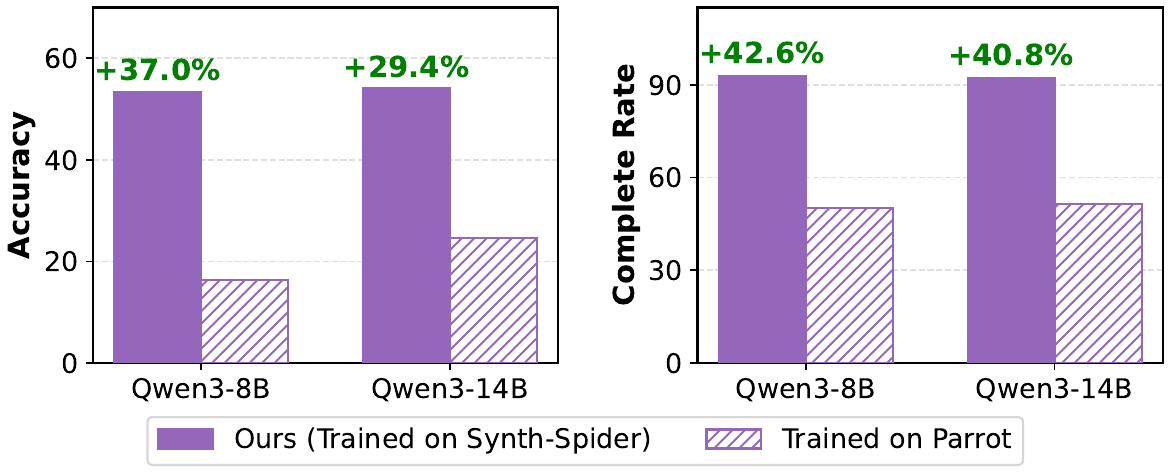}
    \vspace{-0.5em} 
    \caption{Evaluation results on Synth-Bird (Dev) of models trained on Synth-Spider (Train) and Parrot (Train).}
    \label{fig:trained_on_spider_and_parrot_eval_bird}
    \vspace{-0.5em}
\end{figure}
\begin{figure}[t]
    \centering
    \includegraphics[width=0.95\columnwidth]{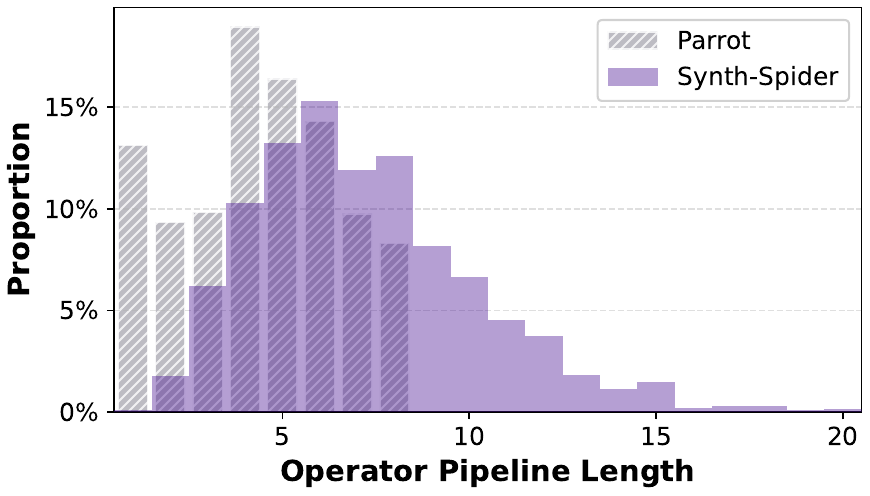}
    \vspace{-1em} 
    \caption{Operator Pipeline Length Distribution.}
    \label{fig:op_len_distribution}
    \vspace{-0.5em}
\end{figure}

\stitle{Exp-\expnum\label{exp:data_effectiveness}: Effectiveness of Synthesized Training Data.}
We evaluate whether models trained on our synthesized Synth-Spider training set generalize better than those trained on the public Parrot training set. Both models are evaluated on the Synth-Bird dev set.

As shown in Figure~\ref{fig:trained_on_spider_and_parrot_eval_bird}, training on Synth-Spider yields much higher performance: accuracy increases by 37.0 and 29.4 points on Qwen3-8B and Qwen3-14B over Parrot. Completion rate shows even larger gains, with increases of 42.6 and 40.8 respectively.
To understand why Synth-Spider yields stronger generalization, we compare operator-pipeline length distributions of the two training datasets in Figure~\ref{fig:op_len_distribution}. Parrot samples cluster on shorter pipelines, while Synth-Spider consistently contains more medium- and long-horizon operator pipelines.
This adds procedural complexity to promotes tree-based agentic reasoning during agentic training, in turn improving generalization capabilities to unseen data.


\begin{shaded}
\noindent \textbf{Finding \findingnum: 
Our synthetic training data provides effective supervision for agentic training, significantly improving accuracy and completion rate on unseen tasks.
}
\end{shaded}
\subsection{Model Scaling Study}
\label{subsec:scalability}

\begin{figure}[t]
    \centering
    \includegraphics[width=1.01\columnwidth]{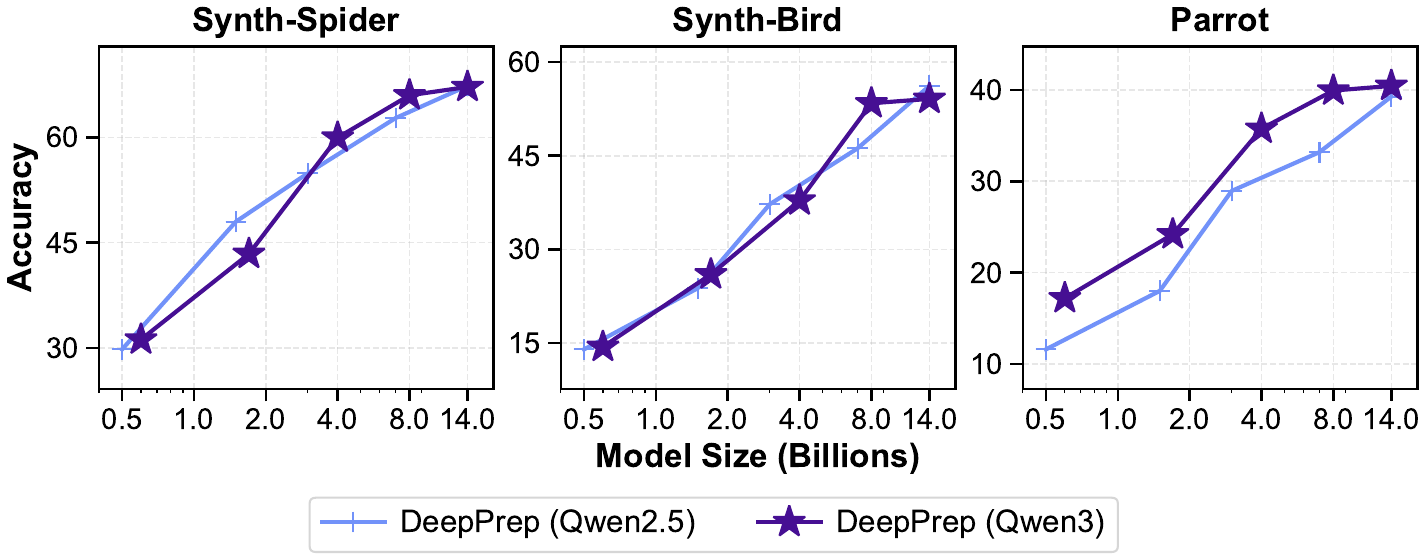}
    \vspace{-1.5em} 
    \caption{Performance scales with larger model size.
    }
    \label{fig:performance_scale_accuracy_only}
    \vspace{-0.5em} 
\end{figure}

\stitle{Exp-\expnum\label{exp:model_scaling}: Model Scaling Study.}
We study how \sys performs under different backbone model sizes. We train \sys with LLM backbones ranging from 0.5B to 14B parameters across the Qwen2.5 and Qwen3 families.
Figure~\ref{fig:performance_scale_accuracy_only} shows that accuracy steadily increases with model size. On Synth-Spider, scaling Qwen3 from 0.6B to 14B improves Accuracy from 31.20 to 67.18. Across comparable model sizes, the Qwen3 series also shows stronger out-of-domain performance than Qwen2.5. For example, on Synth-Bird, Qwen3-8B achieves 53.39 Accuracy versus 46.17 for Qwen2.5-7B. These results suggest \sys benefits from stronger backbone models and maintains positive scaling trends.


\begin{shaded}
\noindent
\textbf{Finding \findingnum: The performance of \sys improves with larger backbone model sizes, which suggest that \sys benefits from stronger backbone models and maintains positive scaling trends.}
\end{shaded}

\section{Related Work}
\label{sec:related_work}

\stitle{Data Preparation.}
Prior data preparation work mainly focuses on programming-by-example (PBE) or specialized sub-tasks. PBE systems~\cite{barowy2015flashrelate, singh2016blinkfill, he2018transform, gulwani2012spreadsheet, jin2017foofah} synthesize transformations from input-output examples, which are effective for well-specified transformations but require example-level supervision for each task.
Other systems target specific data preparation sub-problems~\cite{yang2021auto, fan2024autoprep, lai2025auto, chen2025auto, DBLP:journals/pvldb/LiHYWC23,huang2018auto, fan2024cost}. For example, Auto-Tables~\cite{DBLP:journals/pvldb/LiHYWC23} focuses on table reconstruction operations such as pivoting and stacking. Such systems emphasize structural transformations but are not designed to jointly handle value-level cleaning~\cite{chai2025cost} and multi-table integration~\cite{zhu2017auto}. Text-to-Pipeline~\cite{ge2025text} maps natural language descriptions to pipelines, but assumes detailed step-wise specifications as input.
In contrast, \sys targets end-to-end autonomous data preparation from high-level target schema descriptions and supports a much broader operator space covering cleaning, reshaping, and integration.

\stitle{LLM-Powered Data Preparation.}
Recent work explores using LLMs for data preparation pipeline construction. Prompting-based approaches directly generate transformation code from natural language specifications~\cite{li2024towards,sapkota2025vibe,lai2025auto}. 
To incorporate execution feedback, ReAct-style methods interleave reasoning with tool actions and update subsequent decisions based on observed results~\cite{yao2023react, aksitov2023rest}. Such approaches follow a linear interaction trajectory, where actions are appended sequentially. More recent work explores tree-structured search over reasoning or operator spaces, including Tree-of-Thoughts (ToT)~\cite{yao2023tree} and MCTS-based agents~\cite{li2025alphasql, ge2025monteprep}. For example, MCTS-style approaches typically rely on rollout-based value estimates and scalar reward signals to guide node selection and backpropagation~\cite{hao2023reasoning}. However, data preparation pipelines expose structured execution feedback, such as schema mismatches, execution errors for operators, and intermediate table states. \sys is designed to operate directly over such execution-grounded states, where each node corresponds to materialized tables and branch decisions are guided by structured execution feedback rather than scalar rollout statistics. This design supports operator-level revision and state-aware backtracking during pipeline construction.

\stitle{Data Synthesis for Data Preparation.}
Training data for preparation is limited, especially for tasks requiring multi-step data preparation pipelines. NL2SQL synthesis methods~\cite{li2025omnisql} generate query–result pairs, but typically do not provide operator-level transformation sequences needed for ADP training. Auto-Tables~\cite{DBLP:journals/pvldb/LiHYWC23} proposes a rule-based inverse-operator synthesis strategy for reshaping tasks, mainly targeting structural transformations, which can limit the diversity of synthesized data. Our synthesis framework differs in scope by generating operator pipelines together with source–target table pairs and reversible noise injection, supporting training signals for autonomous data preparation.

\section{Conclusion}
\label{sec:conclusion}
In this paper, we have introduced \sys, an LLM-powered agentic system for autonomous data preparation. To address limitations of existing LLM-powered approaches, \sys organizes data preparation as iterative, execution-grounded interaction with an environment, and introduces tree-based agentic reasoning to support structured exploration and non-local revision based on runtime feedback. \sys also employs a progressive agentic training framework that incrementally builds operator-level competence and tree-based reasoning ability, complemented by data synthesis that provides diverse and complex ADP tasks. Extensive experiments robustly show that \sys consistently outperforms existing open-source baselines and achieves accuracy comparable to strong closed-source models (\eg GPT-5) at substantially lower inference cost. By releasing source code, datasets, and a comprehensive suite of trained models ranging from 0.5B to 14B parameters, \sys offers a practical, cost-efficient, and deployable solution for autonomous data preparation, and openly provides a foundation for future research on execution-grounded agentic systems.


\clearpage
\newpage
\bibliographystyle{ACM-Reference-Format}
\bibliography{citations/ref} 


\end{document}